\documentclass[conference,twocolumn,a4paper]{IEEEtran}
\usepackage{amssymb,amsmath,amsthm} 
\usepackage{cite} 
\usepackage{psfrag}
\usepackage{here}
\usepackage{algorithm}
\usepackage{algpseudocode}
\usepackage[left=1.57cm, right=1.58cm, top=1.71cm,bottom=4.30cm]{geometry}

\usepackage{graphicx,psfrag}
\usepackage{fancyhdr,afterpage,lipsum}
\usepackage{booktabs}
\usepackage{tabularx}
\usepackage{multicol}
\usepackage{multirow}
\usepackage{color}
\usepackage{subcaption}
\usepackage{url} 
\usepackage[latin1]{inputenc} 

\theoremstyle{remark}  

\begin{document}

\title{Privacy Performance of MIMO Dual-Functional Radar-Communications with Internal Adversary}

\author{\IEEEauthorblockN{Isabella W. G. da Silva, Diana P. M. Osorio, and Markku Juntti}
\IEEEauthorblockA{\textit{Centre for Wireless Communications, University of Oulu, P.O.Box 4500, FI-90014, Finland} \\
Emails: \{isabella.wanderleygomesdasilva, diana.moyaosorio, markku.juntti\}@oulu.fi}}

\maketitle
\begin{abstract}
The co-design of radar sensing and communications in dual-functional radar communication systems brings promising advantages for next generation wireless networks by providing gains in terms of the efficient and flexible use of spectrum, reduced costs, and lower energy consumption than in two separate systems. Besides the challenges associated with the conciliation of the conflicting requirements to perform wireless communication and radar sensing in a real-time cooperation, privacy issues represent a cause of concern as the co-design can let the network prone to active attacks. This paper tackles this issue by evaluating the associated privacy risks with the design of transmit precoders that simultaneously optimise both the radar transmit beampattern and the signal-to-interference-plus-noise at the communication users. Our results show that if a malicious user can infer the transmitted precoder matrix with a certain accuracy, there is a reasonable risk of exposure of the location of the target and privacy breaches.
\end{abstract}
\begin{IEEEkeywords}
 dual-functional radar-communications systems, MIMO, precoder design, privacy performance.
\end{IEEEkeywords}
\section{Introduction}
With the growing number of connected devices and the demand of an efficient exploitation of spectral resources, alternative frequency bands and shared spectrum scenarios must be implemented to attend the requirements of future sixth-generation (6G) wireless communications~\cite{art:alwis}. To this end, sharing spectrum resources between communication and radar systems have attracted significant attention as a means to alleviate the spectrum scarcity and under-utilisation. Particularly, integrated sensing and communication (ISAC) has emerged as a new design paradigm and one of the key capabilities of 6G networks that allows for an integration gain by sharing wireless resources and a cooperation gain by the mutual assistance of both functionalities~\cite{art:liu2022}.

In ISAC systems, the dual-functional radar-communication (DFRC) design focuses on the joint design for a real-time cooperation of both functionalities. Particularly, DFRC methods that employ multiple-input-multiple-output (MIMO) schemes provide improved performances by allowing the  simultaneous transmission of probing signals to targets and communication to multiple users, which can be attained by optimisation of transmit beamforming~\cite{art:liu2018,art:yonina}. For instance, the precoder design proposed in~\cite{art:yonina} considers the transmission of individual radar and communication waveforms, which entails on an improved radar transmitted beampattern and satisfy quality of service for the communication users (CUs). Hence, the transmit precoder contains information of the location of targets and data intended to the CUs, which can be inferred and explored by malicious users of the network. In fact, the exposition of sensing information raises huge security and privacy concerns that should be considered for the design of 6G networks and ISAC systems~\cite{art:Wei2022,9097898}. On this matter, security concerns regarding the communication information contained in the radar waveform that can be exposed to untrusted targets have recently gained attention on the literature. For instance, the works in~\cite{art:su2021} and~\cite{art:ren} have focused on the design of secure schemes to prevent eavesdropping from malicious targets. Su \textit{et al.} proposed a beamforming design in~\cite{art:su2021} to minimise the signal-to-noise ratio (SNR) at the eavesdropper target while ensuring a minimal signal-to-interference-plus-noise ratio (SINR) at each CU and beampattern requirements. Moreover, Ren \textit{et al.} considered a network with multiple sensing targets in~\cite{art:ren}, where a portion of them are untrusted, thus a beamforming design was proposed to minimise the beampattern matching error restricted to secrecy rates requirements.

On the other hand, the privacy of the targets' location information on DFRC systems still remains mostly unexplored. To the best of our knowledge, privacy concerns have mainly been evaluated on spectrum sharing between radar and communication systems~\cite{art:Anastasios,art:clark, art:lium}. For instance, in~\cite{art:Anastasios}, Dimas \textit{et al.} considered the case where an adversary hacked the communications system and intends to estimate the position of the radar based on the received precoder matrix. Therein, it was shown that the adversary could estimate the position of the radar with a reasonably high probability, even for different precoding designs. Also, in~\cite{art:clark}, Clark \textit{et al.} assessed the privacy of a primary user in a dynamic spectrum assisted system. Different adversary techniques to exploit the spectrum access system and obfuscation strategies to protect user privacy were evaluated. In~\cite{art:lium},  Liu \textit{et al.} proposed a privacy-preserving mechanism based on a game-theoretic approach to protect primary and secondary users location information.

Recognizing that privacy issues are critical on the design of trustworthy ISAC systems, we intend to fill this gap by investigating a DFRC system with multiple MIMO CUs and a point-like target where one of the CUs is assumed to act as an adversary and tries to infer the location of the target from the received precoder matrix. To this end, the precoder design proposed in~\cite{art:yonina} is extended to the multiuser MIMO case, and it is demonstrated that if the adversary can infer the precoder matrix transmitted by the BS with certain accuracy, a replica of the transmit beampattern can be estimated, thus, the adversary can infer the target's angular position. For the inference problem, a particle filter algorithm  is employed and executed through a number of Monte Carlo simulations. 


\textit{Notation:} Throughout this paper, bold upper-case letters denote matrices whereas bold lower-case letters denote vectors; $(\cdot)^T$ and $(\cdot)^H$ stands for the matrix transpose and Hermitian transpose, respectively; $\mathbf{I}$ is the identity matrix; $||\cdot||$ and $|\cdot|$ are the Euclidean-norm and the absolute value operator; $\mathrm{Tr}(\cdot)$ is the trace of a square matrix; $\Pr(\cdot)$ stands for probability and $p_X$ is the probability density function (PDF) of the variable $X$ ; $\mathbb{E}[\cdot]$ is the expectation operator; and $\mathcal{S}_{N}^{+}$ is the set consisting of all n-dimensional complex positive semidefinite matrices.
\section{System Model}\label{sec:model}
\begin{figure}[h]
    \centering
    \includegraphics[scale=0.4]{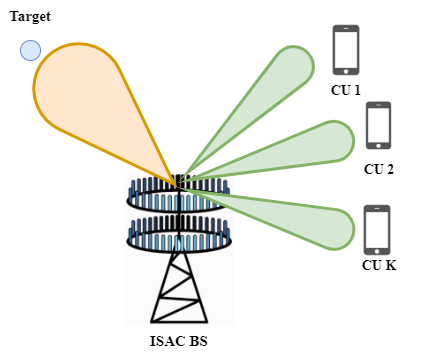}
    \caption{System Model}
    \label{fig:systemmodel}
\end{figure}

Consider the DFRC system illustrated in Fig.~\ref{fig:systemmodel} consisting of a base station (BS) with $M_T$ transmit antennas that intends to communicate to $K$ CUs with $N_R$ receive antennas ($K N_R$$\leq$$ M_T$) while also detecting a point-like target. Similar to~\cite{art:yonina}, we assume that the transmitted signal by the BS is a weighted sum of communications symbols and radar waveforms. Thus, the discrete signal transmitted at a time index $n$ can be written as
\begin{align}
    \mathbf{x}[n] = \mathbf{W}_r\mathbf{s}[n] + \mathbf{W}_c\mathbf{c}[n], n=0,..., N-1,
\end{align}
where $\mathbf{s}[n]$$=$$[s_1[n],..., s_{M_T}[n]]^T$ is the $M_T \times 1$ vector of individual radar waveforms, $\mathbf{W}_r$ is the $M_T \times M_T$ precoder matrix for the radar waveforms, $\mathbf{c}[n]$$=$$[c_1[n],..., c_{K}[n]]^T$ is the vector containing $K \times 1$ parallel communications symbols intended to the $K$ users, and $\mathbf{W}_c$ is the $M_T \times K$ communications precoder matrix.   In addition, the users are assumed to employ a receive beamformer $\mathbf{u}_{k}$, with size $N_R \times 1$, to estimate the transmitted data stream. Accordingly, the estimated data stream and the corresponding signal-to-interference-plus-noise ratio (SINR) at the $k$th user at time $n$ are, respectively, given by
\begin{align}\label{eq:yk}
    y_{k}&= \mathbf{u}_k\left(\mathbf{H}_k\sum_{i=1}^{K} \mathbf{w}_{c,i}c_i + \mathbf{H}_k \sum_{p=1}^{M_T}\mathbf{w}_{r,p}s_p + \mathbf{n}_k\right),\\
    \gamma_k &= \frac{|\mathbf{u}_k^H\mathbf{H}_k\mathbf{w}_{c,k}|^2}{\sum_{i=1 \atop i \neq k}^{K}|\mathbf{u}_k^H\mathbf{H}_k\mathbf{w}_{c,i}|^2+\sum_{p=1}^{M_T}|\mathbf{u}_k^K\mathbf{H}_k\mathbf{w}_{r,p}|^2+ \sigma^2_k||\mathbf{u}_k||^2},
\end{align}
 where $\mathbf{H}_k$ is the $N_R \times M_{t}$ channel coefficient matrix between the BS and the $N_R$ receiving antennas at the $k$th CU, which undergo Rayleigh block fading. $\mathbf{w}_{c,i}$ and $\mathbf{w}_{r,p}$ are the $i$th and $p$th column vectors of $\mathbf{W}_c$ and $\mathbf{W}_r$, respectively. Moreover, $\mathbf{n}_k$ is the noise component at the $k$th user, modelled as signal-independent, zero-mean, additive white
Gaussian noise (AWGN) with variance $\sigma^2_k\mathbf{I}$. 


\section{Precoder Design}
Assuming that the radar receiver has complete knowledge of the transmitted communications waveform, which is also explored for target detection, and under the consideration that the transmit communication waveform is narrow-band and the propagation path is line of sight (LoS), the baseband signal at direction $\theta$ is given by
\begin{align}
    \mathbf{y}[n,\theta] = \mathbf{a}^H(\theta)\mathbf{x}[n],
\end{align}
with $\mathbf{a}(\theta)$ being the array steering vector with direction $\theta$, expressed as
\begin{align}\label{eq:steer}
    \mathbf{a}(\theta)\!\!=\![1 \hspace{2mm} e^{j2\pi \Delta \sin(\theta)} ... \hspace{2mm} e^{j2\pi (M_T-1) \Delta \sin(\theta)}], \in \mathbb{C}^{M_T \times 1},  
\end{align}
where $\Delta$ is the normalized antenna separation. Hence, the correspondening beampattern at direction $\theta$ is given by
\begin{align}
    \mathbf{B}\!=\!\mathbb{E}(|\mathbf{y}[n,\theta]|^2)\!=\! \mathbb{E}(\mathbf{a}^H\!(\!\theta)\mathbf{x}[n]\mathbf{x}^H[n]\mathbf{a}(\theta)\!)\!=\! \mathbf{a}^H\!(\!\theta)\mathbf{R}\mathbf{a}(\theta)\label{eq:beam},
\end{align}
where $\mathbf{R}$ is the covariance matrix of the transmit waveform. Assuming that the communication symbols, $\mathbf{c}[n]$ and the radar waveforms, $\mathbf{s}[n]$ are uncorrelated, $\mathbf{R}$ can be expressed as
\begin{align}
    \mathbf{R}\!=\!\mathbb{E}(\mathbf{x}[n]\mathbf{x}^H[n])\!=\!\mathbf{W}_c\mathbf{W}_c^H\!+\!\mathbf{W}_r\mathbf{W}_r^H. 
\end{align}

Accordingly, we consider the precoder design proposed in~\cite{art:yonina}, which aims to optimise the radar beampattern restricted to transmit power and quality-of-service (QoS) constraints, which can be formulated as
\begin{subequations}
\begin{align}
\mathcal{P}: \min_{\mathbf{W},\mathbf{u}, \alpha} & \frac{1}{L} \sum_{l=1}^L\left|\alpha d\left(\theta_l\right)-\mathbf{a}^H(\theta_l)\mathbf{R}\mathbf{a}(\theta_l)\right|^2\\
\text { s. t. } & \mathbf{R}=\mathbf{W} \mathbf{W}^H \in \mathcal{S}_{M_T}^{+},\label{eq:RW}\\
& {[\mathbf{R}]_{m, m}=\frac{P_t}{{M_T}}, m=1, \ldots,{M_T}, } \\
& \gamma_k \geq \Gamma, k=1, \ldots, K,\label{eq:gammak}
\end{align}
\end{subequations}
where $\mathbf{W}$$=$$[\mathbf{W}_c, \mathbf{W}_r]$, $\alpha$ is a scaling factor, $\{\theta_l\}^{L}_{l=1}$ are sampled angle grids, $d\left(\theta_l\right)$ is the desired beampattern, $P_t$ is the total transmit power, and~\eqref{eq:gammak} is the QoS constraint for the communications users. It imposes that the received SINR at each user must be greater than a given threshold $\Gamma$ to ensure a reliable connection with the BS. 

The considered optimization problem $\mathcal{P}$ is not convex due to the quadratic equality constraint in~\eqref{eq:RW}. Thus, by rewriting $\mathbf{R}$ as 
\begin{align}
    \mathbf{R} &= \sum_{i=1}^{M_T + K} \mathbf{w}_i\mathbf{w}_i^H=\sum_{i=1}^{M_T + K} \mathbf{R}_i,
\end{align}
$\mathcal{P}$ can be addressed by employing a semidefinite relaxation (SDR) strategy, resulting in the following optimization problem
\begin{subequations}
    \begin{align}
\mathcal{P}1: \min_{\substack{\alpha,\mathbf{u},\mathbf{R}\\ \mathbf{R}_1, \ldots, \mathbf{R}_K}} & \sum_{l=1}^L\left|\alpha d\left(\theta_l\right)-\mathbf{a}^H(\theta_l)\mathbf{R}\mathbf{a}(\theta_l)\right|^2, \\
\text {s.t. } &\mathbf{R} \in \mathcal{S}_{M_T}^{+}, \mathbf{R}-\sum_{k=1}^K \mathbf{R}_k \in \mathcal{S}_{M_T}^{+}, \\
& {[\mathbf{R}]_{m, m}=P_t /M_T, m=1, \ldots, M_T,} \\
& \mathbf{R}_k \in \mathcal{S}_{M_T}^{+}, k=1, \ldots, K, \\
&\left(1\!+\!\Gamma^{-1}\right)\!\mathbf{u}_k^H\mathbf{H}_k\mathbf{R}_k \mathbf{H}_k^H \mathbf{u}_k\! \geq\! \mathbf{u}_k\mathbf{H}_k \mathbf{R} \mathbf{H}_k^H \mathbf{u}_k\nonumber\\
&\hspace{4.2cm}\!+\!\sigma_k^2 ||\mathbf{u}_k||^2, \forall k
    \end{align}
\end{subequations}
$\mathcal{P}1$ is solved by considering an alternating optimization approach as described in the follwing steps:

\begin{enumerate}
    \item \textit{Initialization:} Randomly choose  the receive beamformer, $\mathbf{u}_k$ for all $k$.
    \item \textit{Transmit Beamformer update:} Fixing all the receive beamformers, $\mathcal{P}1$ becomes a convex problem and can be efficiently solved with the convex programming toolbox CVX. Given the covariance matrices $\mathbf{R}, \mathbf{R}_1, \ldots, \mathbf{R}_K$ obtained from $\mathcal{P}1$, the precoder matrices $\mathbf{W}_c$ and $\mathbf{W}_r$ can be calculated as 
    \begin{align}
        &\mathbf{w}_k=\left(\mathbf{u}_k^H\mathbf{H}_k \mathbf{R}_k \mathbf{H}_k^H\mathbf{u}_k\right)^{-1/2} \mathbf{R}_k \mathbf{H}_k^H \mathbf{u}_k,\\
        &\mathbf{W}_c = [\mathbf{w}_1, \mathbf{w}_2, ..., \mathbf{w}_K].
    \end{align}
     Whereas $\mathbf{W}_r$ is derived following~\cite[Eq.~33]{art:yonina}.
     \item \textit{Receive Beamformer update:} With the transmit precoder matrices fixed, $\mathbf{u}_k$ is updated using the MMSE receiver as~\cite{7405344}
     \begin{align}
         \mathbf{u}_k\!&=\!\!\left(\!\mathbf{H}_k\!\left(\!\sum_{i=1 \atop i \neq k}^{M_T + K} \mathbf{w}_i\mathbf{w}_i^H\! \right)\!\mathbf{H}_k^H\!+\!\sigma_k^2 \mathbf{I}\right)^{-1}\mathbf{H}_k \mathbf{w}_k, \forall k 
     \end{align}
     Return to step 2 and repeat the process until $||\phi_l-\phi_{l-1}||/||\phi_l||\leq \epsilon$. Where $\phi_l$ is the objective function of $\mathcal{P}1$ calculated at the $l$th iteration of the alternating algorithm, and $\epsilon$ is the convergence threshold.
\end{enumerate}
\section{Adversary Estimation}~\label{sec:Advestimation}
It is assumed that the adversary is capable to extract noisy versions of the precoder matrix $\Tilde{\mathbf{W}}$$=$$[(\mathbf{W}_c + \sigma^2\mathbf{I}), (\mathbf{W}_r + \sigma^2\mathbf{I})]$ from its received signal, and the position of the target, $X$, is considered a random variable, similar to~\cite{art:Anastasios}. Thus, based on the extracted version of the precoder matrix, the adversary tries to estimate the probability distribution of $X$, $p_X$. For this, since the adversary is one of the communications users of the system, we assume that it knows the position of the BS and calculates the angular position of the target in terms of the BS location. This estimation could be treated as a Bayesian inference problem, thus, after $T$ observations, $p_X$ is calculated as 
\begin{align}\label{eq:px}
    p_X(X^1,...,X^T | \Tilde{\mathbf{W}}^1,...,\Tilde{\mathbf{W}}^T)&=\nonumber\\\frac{\Pr(\Tilde{\mathbf{W}}^1,...,\Tilde{\mathbf{W}}^T | X^1,...,X^T)}{\Pr(\Tilde{\mathbf{W}}^1,...,\Tilde{\mathbf{W}}^T)}p_X(X^1,...,X^T).
\end{align} 

Assuming that all possible locations of the target have the same probability and that every observation of the precoder matrix by the adversary are independent among each other, \eqref{eq:px} can be rewritten as
\begin{align}\label{eq:pxnew}
    p_X(X^1,...,X^T | \Tilde{\mathbf{W}}^1,...,\Tilde{\mathbf{W}}^T)&\!=\!\frac{\Pi_{t=1}^T\Pr(\Tilde{\mathbf{W}}^t | X^t)}{\sum \Pi_{t=1}^T\Pr(\Tilde{\mathbf{W}}^t | X^t)}.
\end{align}

Accordingly, the adversary could obtain the optimal estimated distribution for $X$ by calculating \eqref{eq:pxnew} for all possible sequences $\mathbf{X}$$=$$[X^1,...,X^T]$ of the candidate locations. However, for a continuous search area, the number of candidate locations turns the Bayesian inference problem computationally intractable. Hence, Monte Carlo sampling methods can be seen as a good alternative to determine the estimation of $X$, since inference of systems evolving in time can be addressed. One of the most general and simpler Monte Carlo sampling methods is the particle filter algorithm (PFA)~\cite{art:Orhan}. Beyond that, the PFA has a low computational cost, easy implementation, and can be employed on linear or non-linear environments, as well as under Gaussian or non-Gaussian noise~\cite{Fernandez2010}. Under those considerations,  the PFA was considered as a good mean to obtain $p_X$. 

\subsection{Particle Filter Algorithm}

\begin{figure}[t]
    \centering
    \includegraphics[scale=0.36]{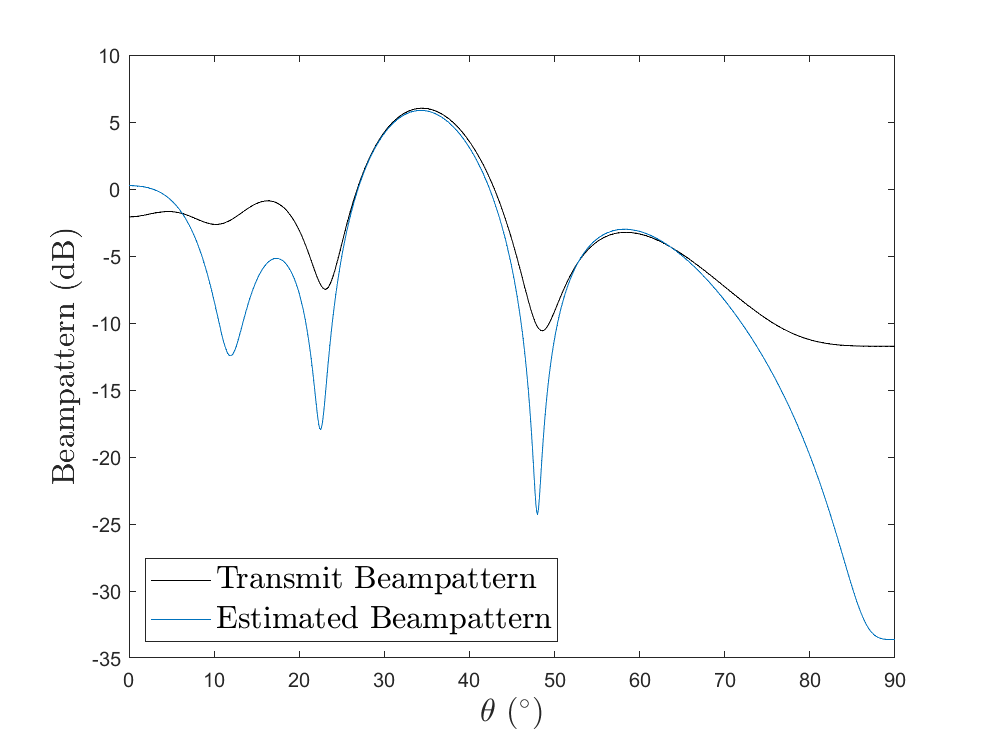}
    \caption{Comparison between the transmit and estimate beam pattern by the adversary.}
    \label{fig:beamcomp}
\end{figure}

For the considered scenario, we assume that all nodes are within a search area, then a set of $M$ particles are randomly selected as possible candidates for the target position at the beginning of the PFA, following a uniform distribution within the considered area, which is defined as $p_{\hat{X}}$. Initially, all particles are assigned equal weights of $\Tilde{q}_i(0) = 1/M$, $i \in \{1,...,M\}$, which are recalculated at each iteration of the algorithm according to 
\begin{align}
     q_i(t) &= \Pr(\mathbf{y}^t|\hat{X}^t_i)\Tilde{q}_i(t-1), i \in \{1,...,M\},\\
     \Tilde{q}_i(t) &= \frac{q_i(t)}{\sum_{i=1}^{M} q_i(t)},
\end{align}
where $\Pr(\mathbf{y}^t|\hat{X}^t_i)$  is the a-priori probability of the adversary observation given the position of the particles. To calculate this probability, we consider that the adversary is able to replicate the steering vector in~\eqref{eq:steer}, and given the noisy version of $\mathbf{W}$, it tries to recreate the transmit beampattern in~\eqref{eq:beam}, i.e., $\mathbf{B}_E$ $=$  $\mathbf{a}^H(\theta)\Tilde{\mathbf{R}}\mathbf{a}(\theta)$, with $\Tilde{\mathbf{R}}$$=$$\Tilde{\mathbf{W}}\Tilde{\mathbf{W}}^H$. As illustrated in Fig.~\ref{fig:beamcomp}, the transmitted beam pattern contains information about the direction of the target with respect to the BS and the noisy version made by the adversary can indeed approach a similar result, which implies that the position of the target is on risk. Also, it is considered a discrete search area consisting of a grid with $N$ cells of equal size, where the midpoint of each cell has its angle and radius calculated in relation to the BS position, i.e., there is  $[\theta_{\mathrm{BS,n}},d_{\mathrm{BS,n}}]$ $\forall$ $n \in [1,N]$. Thus, the estimated beampattern is matched with the corresponding $\theta_{\mathrm{BS,n}}$ of each cell, which entails that for each $\theta_{\mathrm{BS,n}}$ there is a related $\mathbf{B}_E(\theta)$ value, which we refer to as $\mathbf{B}_{En}$. Also, every particle within the $n$th cell share that same value of $\mathbf{B}_{En}$. Finally, since the values of $\mathbf{B}_{En}$ are not restricted between [0,1], we applied a normalisation to employ this result as the probability $\Pr(\mathbf{y}^t|\hat{X}^t_i)$, which can be expressed as
\begin{align}
    \Pr(\mathbf{y}^t|\hat{X}^t_i) = 1-e^{-\mathbf{B}_{En}}, \forall\hspace{1mm}i \in n, n= 1, \ldots, N.
\end{align}

\begin{figure*}[t]
    \centering
    \begin{subfigure}[h]{0.45\textwidth}
        \centering
        \includegraphics[scale=0.4]{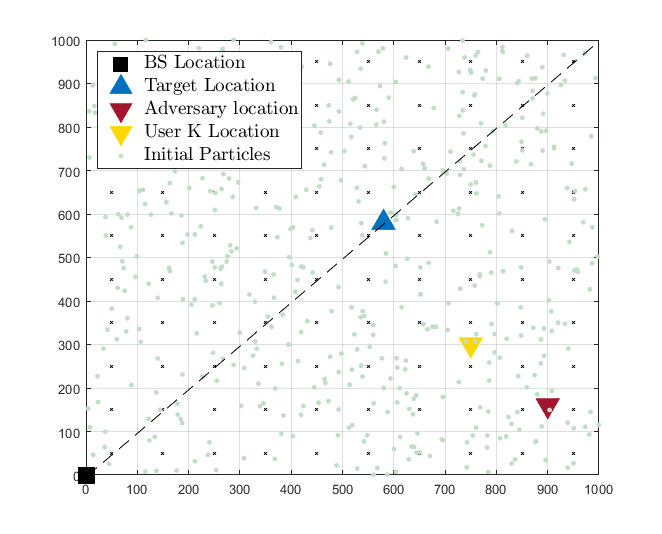}
        \caption{PFA process at $t$$=$$0$.}
    \end{subfigure}%
    \begin{subfigure}[h]{0.5\textwidth}
        \centering
        \includegraphics[scale=0.4]{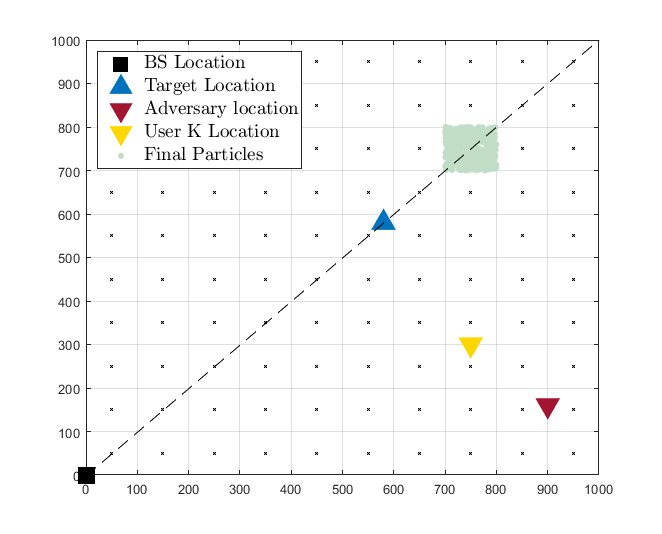}
        \caption{PFA process at $t$$=$$T$.}
    \end{subfigure}
    \caption{Adversary estimation of the target position at $t$$=$$0$ and $t$$=$$T$.}
    \label{fig:grid}
\end{figure*}

Furthermore, to avoid sample depletion, we consider a resampling step for the particle filter algorithm~\cite{art:fredrik}. First, the weighted sum of each cell is calculated and used to define the cells in which the particles will be reallocated to. For this, we adopted a multinomial resampling technique~\cite{art:jol}. First an average weight per cell is calculated based on the weights obtained by the particles in each cell. Then, the purpose of the multinomial resampling is to re-allocate particles with low weights in high-weighted cells. After resampling, all new particles are assigned the same importance weight and the process continues iteratively for each $t \in \{1,...,T\}$ as presented in Algorithm 1. Finally, after all the observations are completed,  the resulting $\hat{X}_i^T$, $i \in \{1,...,M\}$ corresponds to the adversary estimate of the target's angular position. The process followed by the adversary to estimate the angular position of the target is depicted in Fig.~\ref{fig:grid}. At time $t$$=$$0$, the set of $M$ particles is randomly positioned on the grid, and at the time instant $t$$=$$T$, the particles have converged to a cell with a close angular position to that of the target.
 
\begin{algorithm}
\caption{Particle filter algorithm}
\begin{algorithmic}[1]
\footnotesize
       \State Given $\mathbf{y}^1,...,\mathbf{y}^T$, M, $p_{\hat{X}}$, $n_{th}$
       \State $t=0$
        \State Draw $\hat{X}_1^0,...,\hat{X}_M^0$ from $p_{\hat{X}}$
        \State Set $\Tilde{q}_i(0) = 1/M$, $i \in \{1,...,M\}$
        \For{$t$ $\gets$ 1 to $T$}
        \For{$i$ $\gets$ 1 to $M$}
        \State $q_i(t)$ = $\Pr(\mathbf{y}^t|\hat{X}^t_i)\Tilde{q}_i(t-1)$
        \EndFor
        \State $\Tilde{\mathbf{q}}(t) = \mathbf{q}(t)/\sum_{i=1}^{M} q_i(t)$
        \For{$l$ $\gets$ 1 to $N$}
        \State $\Bar{q}_l(t)$ $=$ $\sum_{i=1}^{N}\Tilde{q_i}(t)$
        \EndFor
        \For{$i$ $\gets$ 1 to $M$}
        \State Choose $ind$ from step 11 according to \cite{art:jol}
        \State Draw $\Tilde{X}_i^t$ from $p_{\hat{X}_{ind}}$ 
        \State $\Tilde{q}_i(t)$$=$$ 1/M$, $i \in \{1,...,M\}$
        \EndFor
        \State $\hat{X}_{i}^t$ $=$  $\Tilde{X}_i^t$
        \EndFor
      \State \Return $\hat{X}_i^T$, $i \in \{1,...,M\}$
\end{algorithmic}
\end{algorithm}

\section{Numerical results and discussions}
For the numerical results, the search area is assumed to have 1000 m $\times$ 1000 m  and is divided by cells of 100 m $\times$ 100 m. The BS is assumed to be positioned at the coordinates $(0,0)$, while the target and $K$$=$$2$ users are randomly positioned on the grid. The BS has $M_T$$=$$20$ transmit antennas, and employs an uniform linear array (ULA) with half-wavelength spacing between
adjacent antennas, while the users are equipped with $N_R$$=$$4$ antennas each; the direction grids, $\{\theta_l\}^{L}_{l=1}$, are uniformly sampled from $0^{\circ}$ to $90^{\circ}$ with a resolution of $0.1^{\circ}$; the width of the ideal beam is $\Theta$$=$$10^{\circ}$, thus the desired beampattern is given by
\begin{align}
    d(\theta_l)=\left\{\begin{array}{l}
1, \bar{\theta}_t-\frac{\Theta}{2} \leq \theta_l \leq \bar{\theta}_t+\frac{\Theta}{2}, \\
0, \text { otherwise},
\end{array}\right.
\end{align}
and the path-loss exponent is $\alpha$$=$$3$. Furthermore, unless specified otherwise, the rest of the considered parameters for numerical evaluations are  given as follows: The SINR threshold is set as $\Gamma$$=$$12$ dB, the noise variances of the received signal and of the estimation are respectively given as, $\sigma^2_k$$=$$-100$ dBm and $\sigma^2$$=$$-10$ dBm, the transmit power by the BS is $P_t$$=$$1$,  the convergence threshold for $\mathcal{P}1$ is $\epsilon$$=$$0.01$, and for the particle filter algorithm we assume  1000 Monte Carlo runs with $M$$=$$500$ particles and $T$$=$$1500$ number of observations for each iteration. Moreover, the adversary is considered satisfied with the obtained results if the confidence of estimation, calculated as $\sum_{i=1}^{M}i/M$, $i \in n$, $n$$=$$1,\ldots, N$, is larger than 90\%, i.e., if one of the cells have more than 90\% of the particles on it.
\begin{figure}[t]
    \centering
    \includegraphics[scale=0.4]{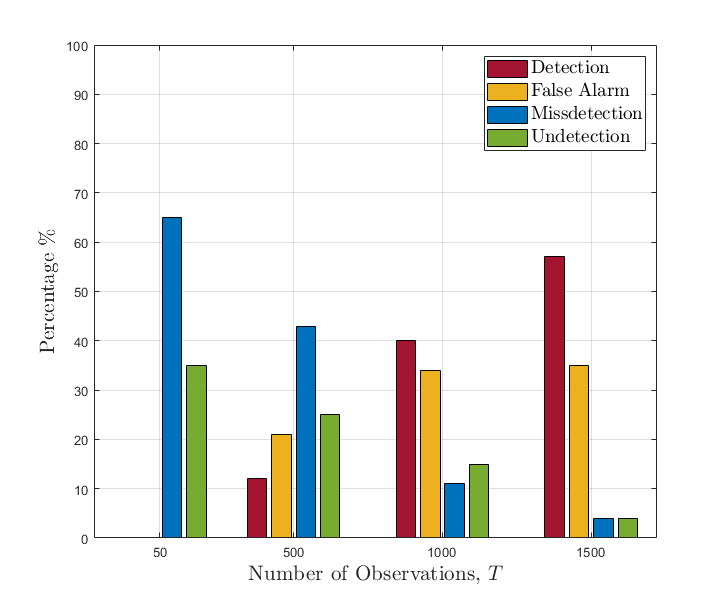}
    \caption{Percentage vs.\ number of observations $T$ for the cases of detection, false alarm, miss detection, and undetection.}
    \label{fig:temp}
\end{figure}

Fig.~\ref{fig:temp} illustrates the percentage vs.\ the number of observations $T$ for different cases obtained at each iteration. In this case, detection is considered to take place, when the absolute angle difference between the real position of the target and the estimated by the adversary is less than $10^{\circ}$ and the confidence is greater than 90\%. For the false alarms, the adversary had a confidence of 90\% or more but did not reach the estimated angle with a difference smaller than $10^{\circ}$ to the real position of the target. Missdetection occurs if the desired confidence was not reached but the estimated angle was indeed within the $10^{\circ}$ difference. Finally, undetection is considered if neither the confidence nor the absolute angle difference is within the imposed requirements by the adversary. Accordingly, note that with a larger number of observations, the confidence of the adversary increases, as expected.  Particularly, the percentage of detections surpasses 50\% for the largest number of observations, thus indicating that if the adversary can recreate the transmitted beampattern based on the received precoder matrix,  it can identify the direction of the target. Besides, note that even with a small number of observations, although the adversary cannot ensure a efficient estimation, it could still identify the target in many cases. 

\begin{figure}[t]
    \centering
    \includegraphics[scale=0.4]{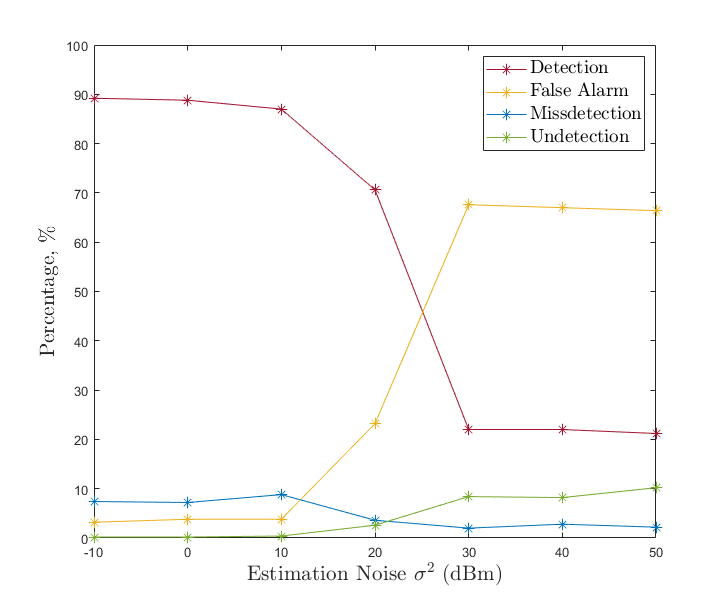}
    \caption{Percentage vs.\ estimation noise, $\sigma^2$ (dBm) for the cases of detection, false alarm, miss detection, and undetection.}
    \label{fig:noise}
\end{figure}

Fig.~\ref{fig:noise} illustrates the percentage vs. the estimation noise $\sigma^2$ of the precoder matrix, for the cases of detection, false alarm, miss detection, and undetection. For this figure, we considered that the adversary is fixed on the coordinates $(800, 100)$, other user is positioned at $(750, 300)$, and the target is fixed at $(550, 400)$. Observe that if the adversary is capable to attain a less noisy version of the precoder matrix, the percentage of detection is significantly high, indicating a high risk of privacy breach. Also, note that for a small estimation noise, there are almost no occurrences of false alarms and undetections, highlighting the risks in terms of privacy preservation if a correct estimation is made by the adversary. On the other hand, as the estimation noise increases, the number of false alarms and undetections significantly increase, although the adversary can still estimate the direction of the target successfully around 20\% of the time, which continues to portrait as a risk for the target's location privacy.

\begin{figure}[t]
    \centering
    \includegraphics[scale=0.48]{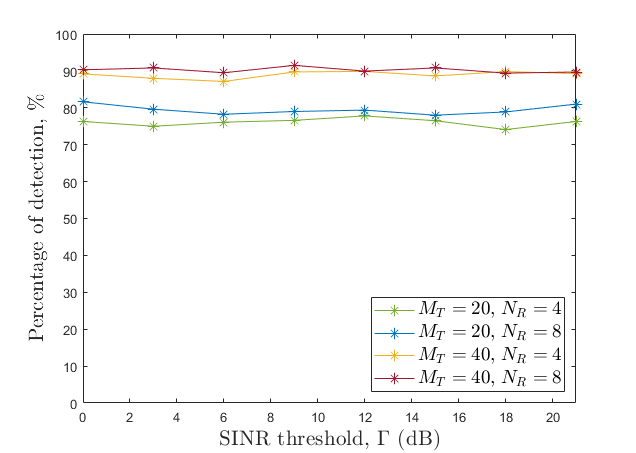}
    \caption{Percentage of detection vs.\ SINR threshold, $\Gamma$ for $M_T$$=$$20$, $40$, and $N_R$$=$$4$, $8$.}
    \label{fig:snrantenas}
\end{figure}

  Fig.~\ref{fig:snrantenas} illustrates the percentage of detection versus the SINR threshold, for different number of transmit antennas at the BS, $M_T$$=$$20,40$, and at the CUs, $N_R$$=$$4, 8$. In this figure, it is assumed the position of the adversary fixed at the coordinates (800, 100), and one more user at (750, 300).  Note that as the number of antennas at the BS increases, the percentage of detection by the adversary also increases. On the other hand, note that the number of receive antennas does not impact significantly on the performance, which is expected since the number of transmit antennas have more influence on the computation of the precoder matrix,  $\mathbf{W}$. In addition, higher values of the SINR threshold does not affect on the estimation accuracy, indicating that the computation of the precoder matrices is robust to different values of $\Gamma$. Accordingly, we can conclude that since both components of the precoder matrix,  $\mathbf{W}_c$ and $\mathbf{W}_r$, and the channel coefficient matrix $\mathbf{H}$ depends on $M_T$, the influence of the number of antennas at the BS on the percentage of detection is more pronounced than $\Gamma$ or $N_R$. 

\section{Conclusions}
In this paper, we investigated the probability of an adversary to estimate the angular position of a target within a DFRC system from noisy versions of the transmit precoding matrix. For the estimation procedure, a particle filter algorithm was employed aiming to identify the direction angle of the target. Also, for the particle filter algorithm, we considered that the a priori probability of the adversary observation is based on a normalisation of the beampattern estimated by the adversary given the noisy versions of the transmitted precoders. The results showed that after a number of observations, the adversary could successfully identify the direction of the target in a reasonable number of cases. Also, this estimation still have means to be improved considering that the adversary have the potential to enhance the accuracy on the estimation of the precoder matrix. 

\section*{Acknowledgement}
This work has been supported by Academy of Finland, 6G Flagship program (Grant 346208) and FAITH Project (Grant 334280).
\bibliographystyle{IEEEtran}
\bibliography{references.bib}
\end{document}